\documentclass[]{revtex4-1}
\usepackage{epsfig}
\usepackage{graphicx}
\usepackage{color} 
\usepackage{amsfonts}
\usepackage{amsmath}

\newcommand{\be}{\begin{equation}}
\newcommand{\ee}{\end{equation}}
\newcommand{\ba}{\begin{eqnarray}}
\newcommand{\ea}{\end{eqnarray}}

\newcommand{\ket}[1]{|#1\rangle}
\newcommand{\de}{\delta}

\begin{document}

\title{Fidelity susceptibility and Loschmidt echo for generic paths 
in a three spin interacting transverse Ising model}
\author{Atanu Rajak}
\email{atanu.rajak@saha.ac.in}
\affiliation{CMP Division, Saha Institute of Nuclear Physics, 1/AF Bidhannagar, Kolkata 700 064, India}
\author{Uma Divakaran}
\email{udiva@iitk.ac.in}
\affiliation{Department of Physics, Indian Institute of Technology, Kanpur 208 016, India}

\begin{abstract}
We study the effect of presence of different types of critical points such as
ordinary critical point, multicritical point and quasicritical point 
along different paths on the Fidelity susceptibility and Loschmidt echo
of a three spin interacting transverse Ising chain using a method which does not involve the language of tensors.
We find that the scaling of fidelity susceptibility and Loschmidt echo with the system size
at these special critical points 
of the model studied, is in agreement with the known results, 
thus supporting our method.
\end{abstract}
\maketitle

\section{Introduction}
The zero temperature quantum phase transitions have been an active area of research for
more than two decades now \cite{sachdev99,chakrabarti96,dutta10}. One of the main motives of these studies is to identify
the quantum critical points (QCP) where the ground state changes significantly
from one form to another and to study the scaling behavior of various quantities around these
critical points. Traditional methods like Landau-Ginzburg theory \cite{chaikin95}
and the relatively new methods of
studying fidelity and fidelity susceptibility \cite{gu10,gritsev09} which are derived from a completely different
area of research, namely, quantum information theory, have provided a lot of insight to the
physics of quantum phase transitions. 

In this paper, we will be focusing on some of the
quantum information theoretic measures like  
fidelity susceptibility $\chi_F$, which
defines the rate at which
fidelity changes in the limit when the two parameters are close to each other.
Here, fidelity is the measure of the overlap of the ground state wavefunction
at two different values of the parameters of the Hamiltonian. 
A dip in fidelity or a peak in the $\chi_F$
as a function of the system parameter signals 
the presence of a phase transition.
These measures show interesting scaling behavior
close to the critical point attracting lots of attention of the scientific community towards it.
For example, the universal scaling 
relation of $\chi_F$ with the system size at the QCP $(\lambda=0)$ and with 
respect to finite but small $\lambda$ is given in terms of some of the 
critical exponents associated with the quantum critical point. It is well established that 
for a $d-$dimensional system of length $L$, the scaling form of $\chi_F$ (see, refs.,\cite{grandi10,albu10,mukherjee11,david09,rams11}) 
at the critical point, say $\lambda=0$, is given by $\chi_F \sim L^{2/\nu-d}$ , 
whereas away from the QCP $(L>>|\lambda|^\nu)$ 
the scaling takes a form $\chi_F\sim |\lambda|^{\nu d -2}$ with $\nu d <2$ \cite{gritsev09}.
For $\nu d>2$, contributions from high energy modes to the fidelity susceptibility
can not be ignored.
However, it has been shown that for some models with $\nu d >2$, fidelity susceptibility
can be used to determine the critical point provided one uses twisted boundary conditions \cite{thakurathi12}. On the other hand,
in the marginal case $\nu d =2$, $\chi_F$ shows logarithmic scaling with $L$
and $\lambda$ \cite{patel13}; 
here $\nu$ is the critical exponent associated
 with the divergence of correlation length at the QCP.
The fact that no previous knowledge about the
order parameter or the symmetry of the system is required to locate the critical points, adds
to the popularity of these measures\cite{chen08}. 
The success of these measures in detecting quantum critical points in a given system is remarkable. 
On the other hand, there are examples of quantum phase transitions which can not be captured
using the general definition of fidelity and fidelity susceptibility \cite{divakaran13,thakurathi12}.

%The studies on quantum phase transitions and their critical behavior
%have been useful in connecting various fields like quantum information theory, 
%quantum computation and condensed matter to each other.
%Infact there is a straightforward connection between quantum information theory and
%those in quantum many body systems.
%Since fidelity describes the sensitivity
%to the dissimilarity between the states, it is related to the measure of loss of information encoded
%in quantum states. On the other hand, $\chi_F$ is intrinsically related to the
%dynamic structure factor of the driving Hamiltonian, i.e., the part of the Hamiltonian
%which is producing quantum fluctuations. For example, starting from the definition
%of fidelity of a thermal state, $\chi_F$ is the thermal fluctuation term such 
%as specific heat for internal energy and magnetic susceptibility 
%for magnetization \cite{you07}. 
%At zero temperature, $\chi_F$ is intrinsically related to
%the ground state energy \cite{chen08}. 

We will be studying one more information theoretic measure in this paper, which
is Loschmidt echo (LE) \cite{quan06,rossini07,damski11,mukherjee12,nag12,sharma12,sharma121,divakaran13}. 
LE is the overlap of two wavefunctions, one is the ground state wavefunction $\ket{\psi_G}$ 
of a Hamiltonian $H(\lambda)$ and evolving as $e^{-iH(\lambda)t}\ket{\psi_G}$, 
and the other is the same state but evolving under slightly different Hamiltonian $H'=H(\lambda+\delta)$.
LE also shows a dip at the QCP, thus enabling its detection. In the language of quantum information theory,
it can be used to detect the quantum to classical transition of a spin-1/2 qubit coupled to a many body system
undergoing a quantum phase transition \cite{quan06,rossini07}.
The notion of LE was actually introduced in connection to
the quantum to classical transition in quantum chaos\cite{peres84,peres93,jalabert01,karku02,cerruti02,cucchietti03} 
and now extended to various other systems undergoing
a QPT like Ising model \cite{quan06}, Bose-Einstein condensate\cite{{zheng08,*zheng09}} and Dicke model \cite{huan09}. 
It has also been studied experimentally using NMR experiments \cite{buchkremer00,zhang09,sanchez09}.

In this paper, we study the scaling of $\chi_F$ and LE along different paths
of a three spin interacting transverse Ising model which consists of
ordinary critical point, multicritical point and quasicritical point.
The method used for this path dependent study is new and simpler than the 
conventional method adopted which involves tensors \cite{venuti07,mukherjee11}.
To the best of our knowledge, none of the studies on $\chi_F$ or LE has considered the path dependence
in as much details as is done in the present paper. 

\section{The Model}
\label{sec_model}
The Hamiltonian of a one-dimensional three-spin interacting Ising 
system of length $L$ in presence of a transverse field $h$ is given by \cite{kopp05,divakaran07}
\be
H= -\frac{1}{2}\sum_{n=1}^{L}\big[\sigma^{z}_n(h+J_3\sigma^x_{n-1}\sigma^x_{n+1})+J_x\sigma^x_n\sigma^x_{n+1}\big]
\label{ham1}
\ee 
where $\sigma^x$ and $\sigma^z$ are the usual Pauli spin matrices,
$J_3$ is the three-spin coupling strength connecting spins at sites
$n, n-1$ and $n+1$, and $J_x$ is the coupling 
constant of the nearest neighbor ferromagnetic interaction 
in x direction. Although the 3-spin interacting term in the Hamiltonian
makes it appear difficult to solve, the above Hamiltonian
can be diagonalized using the standard Jordan-Wigner $(JW)$ transformation \cite{lieb61,pfeuty70,bunder99}
which maps an interacting spin-1/2 system to a system of spinless fermions. 
The Jordan-Wigner transformation relations between spins and fermions are 
defined as 
\ba
&& c_n= \big(\prod_{j=1}^{n-1}\sigma^{z}_j\big)\sigma^{-}_n\nonumber\\
&& \sigma^z_n= 2c^{\dagger}_nc_n-1,
\label{JW}
\ea
where $\sigma^{\pm}_n= (\sigma^x_n \pm \sigma^y_n)/2$, and $c_n$, $c^{\dagger}_n$ are fermionic annihilation 
and creation operators respectively with usual anticommutation 
relations. Substituting the $\sigma-$operators by the JW fermions $c_i$ 
and performing a Fourier transformation, the Hamiltonian (\ref{ham1}) 
takes a form
\ba
H&=& -\sum_{k=0}^{\pi}\big[(h+J_x\cos k-J_3\cos 2k)(c^{\dagger}_kc_k+c^{\dagger}_{-k}c_{-k})\nonumber\\
&+& i(J_x\sin k-J_3 \sin 2k)(c^{\dagger}_kc^{\dagger}_{-k}+c_kc_{-k})\big]\nonumber\\
&=&\sum_k H_k.
\label{ham2}
\ea
The Hamiltonian $H_k$ is a $2\times 2$ matrix when written in a basis 
$\ket{0}$ (with 0 c-fermions) and $\ket{k,-k}$ (=$c_k^{\dagger}c_{-k}^{\dagger} \ket{0}$),
and has a form
\begin{eqnarray} H_k= \left[ \begin{array}{cc}
h+J_x\cos k-J_3 \cos 2k& J_x\sin k-J_3 \sin 2k\\
J_x \sin k-J_3\sin 2k & -(h+J_x\cos k -J_3 \cos 2k)
\end{array} \right].
\label{eq_matrix}
\end{eqnarray}
The above Hamiltonian can be diagonalized after a rotation by an angle $\theta_k/2$ 
which is given by
\ba
\tan \theta_k = \frac{J_x\sin k-J_3 \sin 2k}{h+J_x\cos k -J_3 \cos 2k}
\label{eq_tantheta}
\ea
with the corresponding eigen energy for the $k-th$ mode as
\be
\varepsilon_k= \big(h^2+J_3^2+J_x^2+2hJ_x\cos k- 2hJ_3\cos 2k-2J_xJ_3 \cos k\big)^{1/2}.
\label{energy}
\ee 
The diagonalized Hamiltonian can now be written as
\ba
H=\sum_k \varepsilon_k(\eta_k^{\dagger} \eta_k-1/2)
\label{eq_quasi}
\ea
where $\eta_k$ is the quasiparticle corresponding to the
Hamiltonian $H$.

Now, from Eq.(\ref{energy}) one can easily verify that the low energy 
excitation gap vanishes on the critical lines $h=J_3+J_x$ and $h=J_3-J_x$ 
for the wave vectors 
$k=\pi$ and 0, respectively. These two lines are the critical lines 
separating two phases, the ferromagnetically ordered phase and 
the paramagnetic phase. The long-range 
order in the ferromagnetic phase is present only for a weak transverse 
field lying in the range $J_3-J_x<h<J_3+J_x$.  
The associated quantum critical exponents with these QPTs are the same as in 
the one-dimensional transverse Ising model with $\nu=z=1$, where $\nu$ and $z$
are the correlation length and dynamical exponents, respectively \cite{kopp05}. 
There is also another phase transition at 
$h=-J_3$ between 
three-spin dominated phase and quantum paramagnetic phase. 
This phase transition is analogous to the anisotropic phase transition seen in the one-dimensional transverse XY model. 
%The paramagnetic phase is again divided into a commensurate and an incommensurate region, with incommensurate 
The ordering wave vector $k_0$ in this case is parameter dependent and is given by 
\be
\cos k_0= \frac{J_x(h-J_3)}{4hJ_3}.
\label{inwvector}
\ee  
On the critical line $h=-J_3$, the incommensurate wave vector 
$k_0$ takes a value such that $\cos k_0= J_x/2J_3$
which implies that the anisotropic transition can not occur for $J_3<J_x/2$. 
The two critical lines $h=J_3+J_x$ and $h=-J_3$ meet at a point, called multicritical point (MCP)
and will be the focus of attention in this paper.
Another MCP occurs at the intersection of $h=-J_3$ and $h=J_3-J_x$.
The associated critical exponents with these MCPs is given by $z=2$ and $\nu=1/2$.
\textcolor{blue}{It can be shown that near the multicritical points, there may exist some
special points called quasicritical points, which although do not play a role in determining
the phase diagram of the model, but do affect the scaling of various quantities. This is because, at these quasicritical points,
the energy $\varepsilon_k$ for modes close to the critical mode has a local minima shifted from the critical point.
}

The phase diagram of the model for $J_3=-1$ is shown in Fig. 1.
We study the scaling of $\chi_F$ and LE along different paths in this phase diagram
containing ordinary critical points, multicritical points and quasicritical points,
and in the process develop a scheme which can be extended to study any type of path.
%Our aim is to explore the effect of various types of critical points, namely,
%ordinary critical point, quasi critical point and multicritical point
%on the behavior of $\chi_F$ and Loschmidt echo using a method which
%can be applied to any type of path.
%Our aim is to explore the scaling of $\chi_F$ when the multicritical point is 
%crossed along different paths including a critical line, thereby
%studying the effect of quasicritical modes in their scaling. 
\begin{figure}
\includegraphics[height=2.6in]{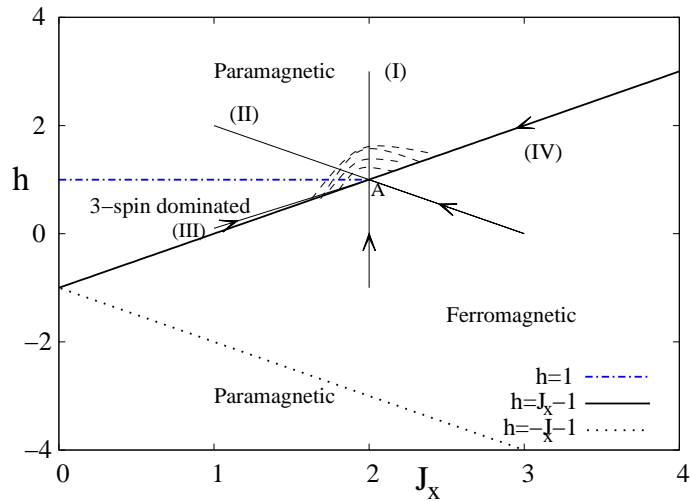}
\caption{Phase diagram of the three-spin interacting transverse Ising model along with the various paths studied for approaching the MCP. 
\textcolor{blue}{The point A corresponds to one of the multicritical points. The phase boundaries are marked by the three different lines
as shown in the label whereas the paths studied in this paper are I, II, III and IV, as shown by the lines with arrows. 
Path IV is also the gapless line separating various phases.
The shaded region corresponds to the region where quasicritical points exists.
}}
\label{fig_phasediagram}
\end{figure}

\section{Fidelity Susceptibility}
\label{sec_chiF}

As defined before, fidelity $(F)$ is defined as the overlap of the ground state 
wavefunction at parameter values separated by a distance 
$\delta$ whereas $\chi_F$ is the rate at which the fidelity changes with
a parameter of the Hamiltonian \cite{gu10,gritsev09,grandi10,albu10,mukherjee11,david09,rams11,thakurathi12,patel13,damski13a,damski13b}.
There are many mathematical forms of
calculating $\chi_F$ \cite{gu10,zanardi06,yang08,gu08}. 
We shall be focusing on one particular form given by
\begin{eqnarray} 
\chi_F=\frac{1}{4L}\sum_k \left(\frac{d\theta_k}{d\lambda}\right)^2
\label{eq_defchi}
\end{eqnarray}
where $\theta_k/2$ is the angle by which the Hamiltonian needs to be rotated
so that it is diagonalized, see discussion around Eq. \ref{eq_tantheta}.
We study the behavior of the fidelity susceptibility along four different
paths, all of them crossing the multicritical point, 
with an approach which does not require the language of tensors as is done in the
previous studies \cite{venuti07,mukherjee11}.  
We chose these paths with specific reasons: Path I and II crosses quasicritical points
along with the multicritical point whereas Path IV is a gapless line which does not have any quasicritical point.
Path III is a special path containing quasicritical points but very close to the critical line 
which might show some interesting behavior.
The Hamiltonian (\ref{ham1}) has three parameters. For convenience we fix $J_3=-1$ and work in
the parameter space spanned by $h$ and $J_x$. We have repeated the calculations
for $J_3=1$ and no major differences are observed.
The paths studied in this paper are shown in Fig. \ref{fig_phasediagram}. 

In our approach to calculating
$\chi_F$ along a path, we rewrite the Hamiltonian in terms of only one variable using
the equation of the path.
We then rotate $H_k$ by an angle $\phi_k$ 
such that the path or the variable that is 
changed, say $\lambda$, is brought to the 
diagonal term. We then evaluate the $\chi_F$ using Eq. \ref{eq_defchi} after 
calculating the angle $\theta_k$. We briefly mention the method of evaluating
the angle $\phi_k$ below. Let $R$ be the rotation matrix with elements $R(1,1)=\cos(\phi_k)=R(2,2)$ and
$R(2,1)=-R(1,2)=\sin(\phi_k)$. Rewriting $H_k$ in terms of only one variable $\lambda$, and 
performing the rotation by an angle $\phi_k$ 
results to a matrix $H_k'=R^T H_k R$. 
The angle $\phi_k$
is then evaluated by demanding that the off-diagonal term in $H_k'$ is $\lambda$ independent.
After substituting for the angle $\phi_k$, the diagonal term in general will have a form 
$a_k \lambda+b_k$ 
and the off-diagonal term is of the form $c_k$.
Let us assume $b_k \sim k^{z_1}$ and $c_k \sim k^{z_2}$ when expanded near the critical mode.
When $z_1<z_2$, the exponent $z$ corresponding to the scaling of $\varepsilon_k$ at the quantum critical point
$\lambda=0$ is equal to $z_1$, i.e., $\varepsilon_k \sim k^{z_1}$ at $\lambda=0$. 
On the other hand, there can arise situations where the path
shows energy minima at $a_k \lambda_0 +b_k=0$ such that 
$\varepsilon_k \sim k^{z_2}$ 
at these special points $\lambda_0$, also called quasicritical points \cite{deng09,mukherjee10}.
It has been shown that it is the exponent $z_2$, different from the actual exponent $z$ at the critical point, 
which will dominate the scaling of various quantities when the quasicritical points exist.
%If the path shows energy minima at $a_k \lambda+b_k=0$ so that
%$\epsilon_k \sim k^3$ (scaling of $c_k$) at $\lambda_0=-b_k/a_k$ and hence varies with $k$, instead of
%$\epsilon_k \sim k^2$ (scaling of $b_k$) at $\lambda=0$ which is the actual critical point 
%where the gap vanishes, then the point $\lambda_0$ is 
%called the quasicritical point. For that particular mode, it will be the scaling at $\lambda_0$
%that will dominate the fidelity susceptibility behavior, although it may or may not be same as at $\lambda=0$. 
%%%In the paths studied here,
%%%the quasicritical point has
%%%$a_k$ independent of $k$, 
%%%$b_k \sim k^2$ 
%%%whereas $c_k \sim k^3$ when expanded near the critical mode,
%%%so that the quasicritical point is defined
%%%by $a_k \lambda+ b_k=0$ at which the energy scales as $k^3$, different from $k^2$ scaling 
%%%observed at the multicritical point (i.e. $\lambda=0$). This condition for the position of
%%%quasicritical point can also be obtained by minimizing the energy for a given mode $k$.
When there is no quasicritical point along the path,
then only $\lambda=0$ will be the 
minimum of the energy, and $\varepsilon_k$ along with other quantities will scale as 
$k^{z}$, $z$ being the minimum of $z_1$ and $z_2$,  as is the case in Path IV.
On the other hand, if $z_1>z_2$, then the dynamics will always be governed by the exponent 
$z_2$ independent of the presence of the quasicritical point.

Below, we present our results on fidelity susceptibility along the four paths crossing the multicritical point
and discuss the effect of presence of quasicritical points in each path. 
Using the definition of $\chi_F$ in Eq. \ref{eq_defchi}, we get
$$\left(\frac{\partial \theta_k}{\partial \lambda}\right)^2=\frac{a_k^2c_k^2}{\varepsilon_k^4}$$
where $\theta_k=\tan^{-1}(c_k/(a_k\lambda+b_k))$.
We shall write explicit expression for $a_k$, $b_k$, $c_k$ and $\phi_k$ obtained by making the
off-diagonal term in the Hamiltonian in Eq. \ref{eq_matrix} independent of $\lambda$ for each path.

\subsection{Path I}

In the first path that we consider, we fix $J_x=2$ and approach the multicritical point $h=1$ by varying $\lambda=h-1$.
In this case, the off-diagonal term in the Hamiltonian $H_k$ in Eq. \ref{eq_matrix} is already $\lambda$ independent, i.e.,
$H_k(1,1)=\lambda+1+2\cos k + \cos  2k=-H_k(2,2)$ and $H_k(1,2)=H_k(2,1)=2\sin k+\sin 2k$.
Expanding around the critical mode $k_c=\pi$, we get $a_k \sim 1, b_k \approx -k^2$ and $c_k \approx -k^3$.
With $\theta_k=\tan^{-1}(H_k(1,2)/H_k(1,1))$, the susceptibility is given by

\begin{eqnarray}
\chi_F&=&\frac{1}{4L}\sum_{k>0}\big(\frac{\partial \theta_k}{\partial \lambda}\big)^2
=\frac{1}{4L}\sum_{k>0}\frac{(2\sin k+\sin 2k)^2}{\varepsilon_k^4}\nonumber\\
&\approx&\frac{1}{4L}\sum_{k>0}\frac{k^6}{\varepsilon_k^4}\nonumber
\end{eqnarray}
which gives rise to $L^5$ scaling as $\varepsilon_k \sim k^3$ at the quasicritical point $\lambda\sim k^2$. 
Here, we have redefined $(k-k_c)$ as $k$ and expanded $a_k$, $b_k$ and $c_k$ around $k \rightarrow 0$
which will be followed throughout the paper. Note that here $z_1=2$ and $z_2=3$ such that $z=z_1$ but the
scaling of $\chi_F$ is dictated by $z_2$.

\begin{figure}
\includegraphics[height=2.5in]{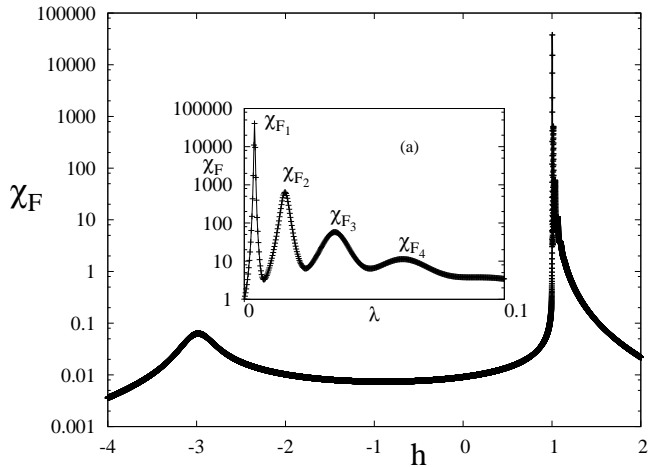}
\caption{Variation of $\chi_F$ as a function of h at $J_x=2, J_3=-1$ 
for a system size L=100. The first peak at $h=-3$
corresponds to the Ising critical point showing linear scaling with $L$
 and the second peak is at the MCP, i.e., at h=1
where the $L^5$ scaling is observed. Inset shows the oscillating fidelity susceptibility close to
the multicritical point pointing to the presence of quasicritical points.}
\label{fig_chih}
\end{figure}

\begin{figure}
\includegraphics[height=2.5in]{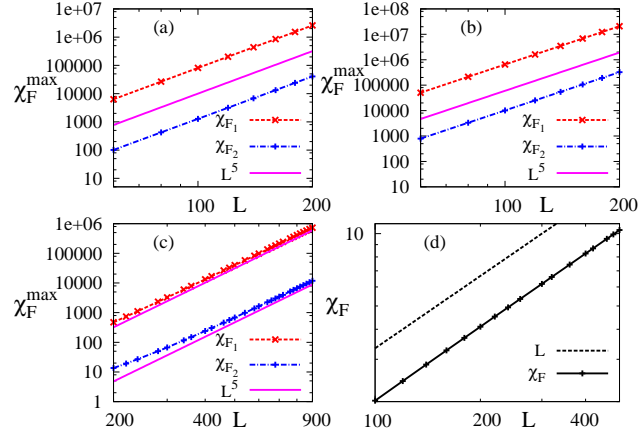}
\caption{Scaling of $\chi_F$ along four different paths:(a), (b) and (c)
corresponds to Path I, II and III with $\chi_F~ \propto ~L^5$ showing the effect of multicritical
point whereas Path IV is linear in $L$ as there is no quasicritical point along this line or path.}
\label{fig_scalingchi}
\end{figure}

The variation of $\chi_F$ as a function of $h$ is shown in Fig. 
\ref{fig_chih} whereas its behavior close to the multicritical point 
$h=1$ shows oscillations as is shown in the inset of the same figure. 
These oscillations can be explained as follows: the quasicritical point occurs
when $\lambda=-b_k/a_k$. Since momentum $k$ is quantized in units of $2\pi/L$, all the allowed momentums near the
critical mode $k_c$, i.e. $k=k_c+2\pi m/L$ for integer m,  will also show $\varepsilon_k \sim k^3$ behavior. Each value of $k$
will give rise to a different value of $\lambda$ close to $\lambda=0$ resulting to more than one 
quasicritical point near the multicritical point \cite{mukherjee11}.
The scaling with of $\chi_F$ with $L$ along this path is shown in Fig. \ref{fig_scalingchi}a for the first two peaks
occurring in $\chi_F-\lambda$ plot (see inset of Fig. \ref{fig_chih}).

\subsection{Path II}
%\label{b}

We now consider the path $h+J_x=3$, path II in Fig. 1  and approach the MCP varying $\lambda=h-1$. 
After performing a rotation by an angle $\phi_k$ to bring $\lambda$ to the diagonal term, we get

%\ba
% H_k(11)&=&\lambda\big[\cos 2\phi_k-\cos (k-2\phi_k)\big]\nonumber\\
%&+&\big[\cos 2\phi_k+2\cos (k-2\phi_k)+\cos (2k-2\phi_k)\big]\nonumber\\
%&=&\lambda a+b\nonumber\\
%H_k(12)&=&-\sin 2\phi_k+2\sin(k-2\phi_k)+\sin(2k-2\phi_k)=c\nonumber\\
%{\rm where} \tan(2\phi_k)&=&\frac{\sin k}{-1+\cos k}
%\label{ham2}
%\ea

\ba
 a_k&=&\cos 2\phi_k-\cos (k-2\phi_k) \nonumber\\
b_k&=&\cos 2\phi_k+2\cos (k-2\phi_k)+\cos (2k-2\phi_k)\nonumber\\
c_k&=&-\sin 2\phi_k+2\sin(k-2\phi_k)+\sin(2k-2\phi_k)\nonumber\\
{\rm and}\tan(2\phi_k)&=&\frac{\sin k}{-1+\cos k}.
\ea

When expanded around the critical mode, $a_k\approx 2$, $b_k\approx -k^2$, $c_k\approx -k^3/2$ resulting to
a quasicritical point at $\lambda=k^2/2$ where $\varepsilon_k \sim k^3$.
Since $\theta_k=\tan^{-1}(H_k(1,2)/H_k(1,1))$, $\chi_F$ is given by
\be
\chi_F=\frac{1}{4L}\sum_{k>0}\frac{a_k^2c_k^2}{\varepsilon_k^4}\approx\frac{1}{4L}\sum_{k>0}\frac{k^6}{\varepsilon_k^4}
\ee
This once again results to $L^5$ scaling as 
also confirmed numerically in Fig. \ref{fig_scalingchi}b. 

\subsection{Path III}
%\label{c}
The Hamiltonian $H_k$ after rotation by an angle $\phi_k$
along the path $h-0.9J_x=-0.8$
has the following elements:

%\ba
% H_k(11)&=&\lambda\big[\cos 2\phi_k+\frac{10}{9}\cos (k-2\phi_k)\big]\nonumber\\
%&+&\big[\cos 2\phi_k+2\cos (k-2\phi_k)+\cos (2k-2\phi_k)\big]\nonumber\\
%&=&\lambda a +b\nonumber\\
%H_k(12)&=&-\sin 2\phi_k+2\sin(k-2\phi_k)+\sin(2k-2\phi_k)=c\nonumber\\
%{\rm with} &\tan(2\phi_k)&=\frac{10/9\sin k}{1+10/9\cos k}
%\label{ham3}
%\ea

\ba
 a_k&=&\cos 2\phi_k+\frac{10}{9}\cos (k-2\phi_k)\nonumber\\
b_k&=&\cos 2\phi_k+2\cos (k-2\phi_k)+\cos (2k-2\phi_k)\nonumber\\
c_k&=&-\sin 2\phi_k+2\sin(k-2\phi_k)+\sin(2k-2\phi_k)\nonumber\\
{\rm with} &\tan(2\phi_k)&=\frac{10/9\sin k}{1+10/9\cos k}.
\label{ham3}
\ea
%This results to
%\be
%\chi_F=\frac{1}{4L}\sum_{k>0}\frac{a_k^2c_k^2}{\varepsilon_k^4}\approx\frac{1}{4L}\sum_{k>0}\frac{k^6}{\varepsilon_k^4}
%\ee
After expanding close to the critical mode $k_c=\pi$, we get $a_k\approx-1/9$, $b_k\approx-k^2$ and $c_k\approx9k^3$
with quasicritical point at $\lambda=-9k^2$.
Since $\varepsilon_k \sim k^3$ at the quasicritical point, $\chi_F \sim L^5$, also shown in Fig. \ref{fig_scalingchi}c.
Along all the above three paths, quasicritical points exist either in the paramagnetic phase or in the
three spin dominated phase of the system close to MCP.
It can be seen in Fig. \ref{fig_phasediagram} that path III is very close to the critical line
$h=J_x-1$, which as discussed in the next sub-section, does not have any quasi critical point.
We chose this path to check the effect of this proximity on the scaling behavior
of fidelity susceptibility. Although we do observe $L^5$ scaling, but only for large $L$ and the small deviation for smaller
$L$ which is not seen in Paths I and II could be due to its proximity to the critical line.
Let us try to explore this path further.
The quasicritical point in this path exist at $\lambda=-9k^2$ where $k$ is inversely proportional to $L$.
The factor of $9$ compared to $1$ in path I and $1/2$ in path II shifts the location of quasicritical 
point farther away from the actual critical point. Since we expanded $a_k$, $b_k$ and $c_k$ around the
critical mode and the critical point which may not be correct in this path for small $L$,
we observe a deviation from $L^5$ scaling for small $L$.

\subsection{Path IV}
\label{d}
We finally consider the critical line $h-J_x=-1$ 
with $\lambda=h-1$. Performing rotation by an angle $\phi_k$
to make the off-diagonal term of $H_k$ in Eq. \ref{eq_matrix}
$\lambda$ independent, we get the following form 
of functions for $a_k$, $b_k$, $c_k$ and $\phi_k$:

\ba
 a_k&=&\big[\cos 2\phi_k+\cos (k-2\phi_k)\big],\nonumber\\
b_k&=&\big[\cos 2\phi_k+2\cos (k-2\phi_k)+\cos (2k-2\phi_k)\big],\nonumber\\
c_k&=&-\sin 2\phi_k+2\sin(k-2\phi_k)+\sin(2k-2\phi_k), \nonumber\\
{\rm and }\tan(2\phi_k)&=&\frac{\sin k}{1+\cos k},
\label{ham4}
\ea
which when expanded around the critical mode $k_c=\pi$ gives $a_k\approx k$, $b_k\approx k^3$ and $c_k\approx -k^2$.
We note that for paths I, II and III, $a_k$'s are independent of $k$ and we get quasicritical points 
with minimum energy. But for path IV, $a_k$ is k-dependent and its exponent is less than $z_1$ so that we can ignore the term 
$b_k$ for non-zero $\lambda$. Thus, for all
non-zero $\lambda$, $\varepsilon_k$ goes as $k$, i.e, path IV is critical line,
and we can not get any quasicritical point near the MCP
for which energy is minimum.
Since $\varepsilon_k \sim k^2$ at the MCP $\lambda=0$ which is also the 
dominant point, 
$\chi_F$ scales linearly with $L$, also confirmed numerically in Fig. \ref{fig_scalingchi}d.

\section{Loschmidt echo (LE)}
\label{sec_LE}

As mentioned before, LE is defined as the overlap between two states
differing from each other in the Hamiltonian with which they
are evolving but both starting from the ground state of one of the Hamiltonians \cite{quan06,sharma12,sharma121}. 
Mathematically, if $\ket{\psi_G}$ is the ground state of Hamiltonian $H(\lambda)$ with energy $E_g$, then the LE or $\mathcal{L}$ is given by
\ba
\mathcal{L}(\lambda,t)=\big| \langle\psi(\lambda+\delta,t)|\psi(\lambda,t)\rangle \big|^2
=\big| \langle\psi(\lambda+\delta,t)|\psi_G\rangle\big|^2\nonumber \\
\ea
where $\ket{\psi(\lambda,t)}=e^{-iH(\lambda)t}\ket{\psi_G}=e^{-iE_gt}\ket{\psi_G}$
and $\ket{\psi(\lambda+\delta,t)}=e^{-iH(\lambda+\delta)t}\ket{\psi_G}$,
and $t$ corresponds to time. 
%\textcolor{blue}{Here, $H(\lambda+\delta)$ is the composite Hamiltonian
%of Hamiltonian in Eq.\ref{ham1} and a coupling of a qubit with a term 
%(in this case $\lambda$) of the Hamiltonian of interaction strength $\de$.}
It is easier to calculate LE in the momentum
representation by noting the fact that the Hamiltonian is 
decoupled in the momentum space and hence the ground state wavefunction
can be written as
\ba
\ket{\psi_G}=\prod_k \ket{\phi_k}=\prod_k \cos(\theta_k^{\lambda}/2) \ket{0}+\sin(\theta_k^{\lambda}/2)\ket{k,-k}\nonumber \\
\ea
where $\ket{\phi_k}$ is the ground state of $H_k(\lambda)$ and $\theta_k^{\lambda}/2$ as before,
is the angle by which $H_k(\lambda)$ needs to be rotated to diagonalize it. 
Thus,
$$\mathcal{L}(\lambda,t)=\prod_k \mathcal{L}_k(\lambda,t)=\prod_k \big |\langle \phi_k|e^{iH_k(\lambda+\delta) t}|\phi_k \rangle\big |^2.$$
To calculate the above expression, it is to be noted that $\ket{\phi_k}$ is not an eigenstate of
$H_k(\lambda+\delta)$. Therefore, one needs to find an expression of $\ket{\phi_k}$ in terms of 
eigenstates of $H_k(\lambda+\delta)$ which we denote as $\ket{1}$ and $\ket{2}$.
It can be shown that
\ba
\ket{\phi_k}=\cos \alpha_k \ket{1} + \sin \alpha_k \ket{2}
\ea
where $2\alpha_k=\theta_k^{\lambda}-\theta_k^{\lambda+\delta}$. 
Substituting this form of $\ket{\phi_k}$ in the expression of LE, we get
\ba
\mathcal{L}(\lambda,t)=\prod_k \mathcal{L}_k(\lambda,t)=\prod_k \big(1-\sin^22\alpha_k \sin^2(\varepsilon_k(\lambda+\de)t) \big).\nonumber \\
\label{eq_le1}
\ea
We shall be using the above expression for the calculation of LE
taking into account the effect of path in $\alpha_k$ and $\varepsilon_k$.
For analyzing the behavior of LE, we define a partial sum $S=\ln \mathcal{L}$
along the lines similar to Ref. \cite{quan06}.

\begin{figure}
\includegraphics[height=2.1in]{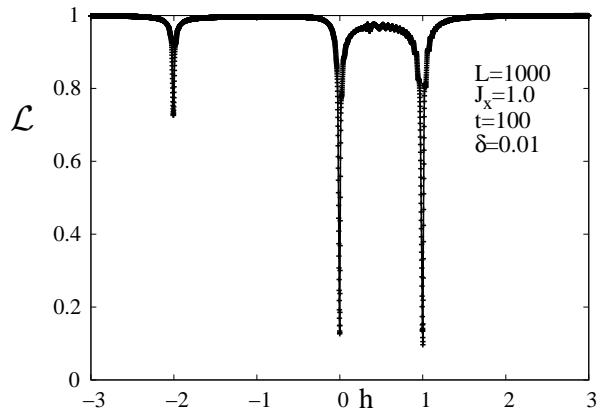}
\caption{LE shows sharp dips at the Ising critical points (h=-2,0)
and at anisotropic critical point at h=1.}
\label{fig_legeneral}
\end{figure}

We first demonstrate the applicability of LE as a tool to detect
the presence of a critical point by taking a path parallel to 
path I of the previous section at $J_x=1$ 
which has three critical points as $h$ is varied for a fixed time $t$. 
Fig. \ref{fig_legeneral} shows that LE can successfully detect all the
critical points in its path. 
It was shown in Ref \cite{quan06} that at the critical point, LE shows decay and revival as a 
function of time which is an indicator of the presence of critical point. The time period
of oscillations is proportional to $L$ in case of Ising critical point
but can vary in a non-linear way for other types of critical points \cite{sharma12}.
We shall demonstrate the difference in the LE
behavior between the anisotropic critical point (or any critical point), 
the multicritical point, and also the quasicritical point 
by studying the behavior of short time decay and 
time period of LE oscillations \cite{quan06}
for the three spin interacting model discussed in section \ref{sec_model}.
In this case also, we use the same method of including the effect of path
as discussed in section \ref{sec_chiF}.

\subsection{Anisotropic Critical Point (ACP)}
\label{a}
\begin{figure}
\includegraphics[height=2.4in]{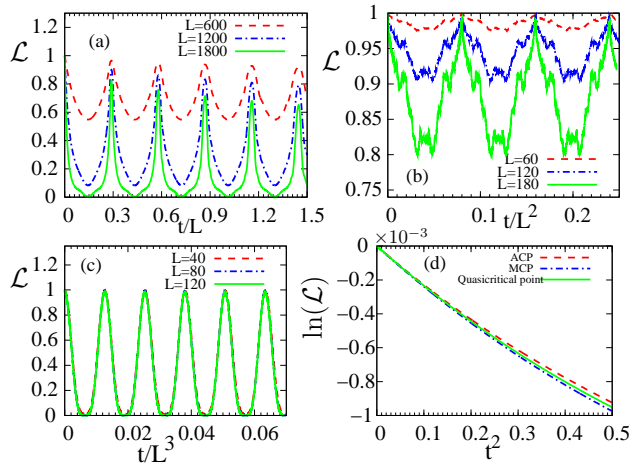}
\caption{\textcolor{blue}{Variation of LE as a function of scaled time $t/L^{\alpha}$ to highlight
the scaling of timeperiod with system size $L$, with 
$\alpha$ being the scaling exponent ($T \propto L^{\alpha}$) 
as obtained in the text for various types of critical points, i.e.,
(a) the anisotropic critical point where $T \propto L$,
(b) multicritical point where $T \propto L^2$ and 
(c) quasicritical point with $T \propto L^3$.
In (d), we present the almost linear variation of $\ln \mathcal{L}$ with $t^2$ for small times
with $L=100$ at the various critical points confirming the general small time behavior given by $\mathcal {L} \sim e^{-\Gamma t^2}$.}}
\label{fig_le2}
\end{figure}

{\it Short time decay:}
On the ACP line (see discussion around Eq. \ref{inwvector}), energy gap vanishes for the critical mode $k_c=\cos^{-1}(\frac{-J_x}{2})$. 
Now, to study the short time behavior of LE, we fix $J_x=1$ and change $h$.
Note that the transverse field $h$ in the Hamiltonian $H_k$ is already in the diagonal term
and hence need not be rotated
similar to path I in section \ref{sec_chiF}. 
We now expand Eq.(\ref{eq_le1}) around the critical
mode  $k_c=\cos^{-1}(-1/2)$
to obtain  $\sin^2\varepsilon_k(h+\delta)t \approx \left(h+\delta-1\right)^2t^2$ 
and $\sin^2\left(2\alpha_k\right)\approx 9k^2\de^2/4(h-1)^2(h+\de-1)^2$.
For small time $t$, this results to $S \propto -\Gamma t^2$, i.e.,
\be
\mathcal{L}\left(h,t\right)\approx \exp\left(-\Gamma t^2\right)
\label{lc2} 
\ee
where the decay constant $\Gamma\propto \de^2/(h-1)^2L^2$. 
Using the expression of $\mathcal{L}\left(h,t\right)$, one can easily show that it remains invariant 
under the transformation $L\rightarrow L \alpha$ and $t\rightarrow t\alpha$ for fixed $\de$, $\alpha$ being some integer.
These scaling relations are also confirmed by the collapse and revival of LE (see Fig.~\ref{fig_le2}a). 
%For short time decay if we double the system 
%size then same amount of decay of LE take double time than earlier.

{\it Time Period analysis}:
Collapse and revival has been seen setting
the parameter values $h=1-\de$,$J_x=1.0$ and $\de=0.01$. 
We can expand $\varepsilon_k(h+\de)$ close to critical mode $k_c$ which gives
$\varepsilon_k(h+\de)\approx \sqrt{4-J_x^2}(k-k_c)$. The dominant contribution 
to $\mathcal{L}(h,t)$ comes from the mode $k=k_c+2\pi/L$ in the limit of large $L$.
One can see from the expression of LE in Eq. \ref{eq_le1} that
the time dependence comes from the term $\sin^2(\varepsilon_k(h+\de)t)$. 
Therefore, the quasi period of this collapse and revival is given by
\be
T=\frac{L}{\sqrt{4-J_x^2}}.
\ee
which is presented in Fig. \ref{fig_le2}a for three different system sizes.

\subsection{Multicritical Point (MCP)}
%\label{b}
In this case we consider the critical line $h-J_x=-1$ 
and approach the MCP ($J_x=2$ and $h=1$) by changing $\lambda=h-1$. 
This is identical to the Path IV studied in section \ref{sec_chiF}.
This path is chosen to study the effect of absence of QCPs on the LE.
We perform a similar rotation of the Hamiltonian in Eq. \ref{eq_matrix}
as done in the case of path IV to shift the $\lambda-$ dependence solely
to the diagonal term.
{\it Short time decay}:
Expanding around the critical mode 
$k_c=\pi$ and assuming short time, we get  $\sin^2\varepsilon_k(\lambda+\delta)t \approx \left(\lambda+\de\right)^2k^2t^2$ and
$\sin^2\left(2\alpha_k\right)\approx k^2\de^2/\lambda^2(\lambda+\de)^2$
resulting to an exponential decay of LE, also observed in the Anisotropic critical point
studied in Part A of this section. In this case, $\Gamma\propto \de^2/\lambda^2L^4$ so that $\mathcal{L}\left(\lambda,t\right)$ is invariant under the 
transformation $L\rightarrow L \alpha$ and $t\rightarrow t\alpha^2$ with fixed $\de$. Again these scalings can be verified by using the collapse and revival 
of LE as a function of time (see Fig.~\ref{fig_le2}b).
%\be
%LE\left(\lambda,t\right)\approx \exp\left(-\Gamma t^2\right)
%\label{lc3} 
%\ee
%where, $\Gamma$ is a function of parameters $\lambda$, $\delta$ and the system
%size $L$.
 
{\it Time Period analysis}
The collapse and revival of LE as a function of time at the MCP can be seen
by setting $\lambda=-\de$ and $\de=0.01$ (path IV ($h-J_x=-1$)). At this point,
$\varepsilon_k(\lambda+\de)\approx 4\pi^2/L^2$, 
%$\sin \theta_k(\lambda+\de)\approx 1$, 
%$\cos \theta_k(\lambda+\de)\approx 1$,
%$\sin \theta_k(\lambda)\approx 0$, and $\cos \theta_k(\lambda)\approx -1$. 
%Then $\sin^2(2\alpha_k)\approx\cos^2\theta_k(\lambda)\approx 1$ and 
which gives time period of oscillation T as
\be
T=\frac{L^2}{2\pi}.
\ee
This is also verified numerically in Fig. \ref{fig_le2} b.

\subsection{Quasicritical Point}
%\label{c}
{\it Short time decay}:
We once again use the path I of section \ref{sec_chiF} with $J_x=2$
so that the path contains quasicritical points in addition to multicritical point as discussed in section \ref{sec_chiF}.
Expanding around the critical mode 
$k_c=\pi$ and assuming short time, we get  $\sin^2\varepsilon_k(h+\delta)t \approx \left(h+\de-1\right)^2t^2$ and
$\sin^2\left(2\alpha_k\right)\approx k^6\de^2/(h-1)^2(h+\de-1)^2$. 
Once again 
\be
\mathcal{L}\left(\lambda,t\right)\approx \exp\left(-\Gamma t^2\right)
\label{lc3} 
\ee
where, $\Gamma\propto \de^2/(h-1)^2L^6$. With
$L\rightarrow L \alpha$, $t\rightarrow t\alpha^3$ and fixed $\de$, Eq.\ref{lc3} remains invariant. 

{\it Time Period analysis:}
The collapse and revival of LE as a function of time at a quasicritical point is obtained
for $h=1-\de+4\pi^2/L^2$, $J_x=2.0$ and $\de=0.01$ (path I) where
$\varepsilon_k(h+\de)\approx 8\pi^3/L^3$.
%$\sin \theta_k(h+\de)\approx 1$, 
%$\cos \theta_k(h+\de)\approx 1$,
%$\sin \theta_k(h)\approx 0$, and $\cos \theta_k(h)\approx -1$. 
%Then $\sin^2(2\alpha_k)\approx\cos^2\theta_k(h)\approx 1$ and 
The time period T of oscillation is then given by
\be
T=\frac{L^3}{4\pi^2}
\ee
as also confirmed numerically in Fig. \ref{fig_le2} c.

\section{Conclusions}
In this paper, we have proposed a method which can be used to study fidelity susceptibility and Loschmidt Echo
for a generic path and verified our method by studying a three site interacting Transverse Ising model. Using this method,
we studied the scaling of fidelity susceptibility and Loschmidt echo with the system size along different paths
which consists of ordinary critical point, quasicritical point and multicritical point. We discuss in details how the scaling
changes due to the presence and absence of quasicritical points.
We also studied the system size dependence of time period of oscillations in the case of Loschmidt echo at
critical point, multicritical point and quasicritical point for different paths. 
To the best of our knowledge, there has not been any study
in such details as done here, involving the effect of various paths on $\chi_F$ and $LE$.

\begin{center}
\bf{Acknowledgements}
\end{center}
The authors acknowledge Amit Dutta for critically reading the manuscript. AR is grateful to Shraddha Sharma for valuable comments and thanks IIT Kanpur for hospitality during this work.
One of the authors (UD) acknowledges funding from the INSPIRE faculty fellowship by the Department of Science and Technology, Govt. of India.

\bibliographystyle{unsrt}
\bibliography{3spin}

\begin{thebibliography}{10}

\bibitem{sachdev99}
S.~Sachdev.
\newblock {\em Quantum Phase Transitions}.
\newblock Cambridge University Press, 1999.

\bibitem{chakrabarti96}
B.~K. Chakrabarti, A.~Dutta, and P.~Sen.
\newblock {\em Quantum Ising phases and transitions in transverse Ising
  models}.
\newblock Springer Heidelberg, 1996.

\bibitem{dutta10}
A.~Dutta, U.~Divakarn, D.~Sen, B.K. Chakrabarti, T.F. Rosenbaum, and G.~Aeppli.
\newblock arXiv:1012.0653, 2010.

\bibitem{chaikin95}
P.~M. Chaikin and T.~C. Lubensky.
\newblock {\em Principles of Condensed Matter Physics}.
\newblock Cambridge University Press, 1995.

\bibitem{gu10}
Shi-Jian Gu.
\newblock {\em Int. J. Mod. Phys. B}, 24:4371, 2010.

\bibitem{gritsev09}
V.~Gritsev and A.~Polkovnikov.
\newblock In L.~D. Carr, editor, {\em Understanding Quantum Phase Transitions}.
  Taylor and Francis, Boca Raton, 2010.
\newblock arxiv:0910.3692.

\bibitem{grandi10}
C.~De Grandi, V.~Gritsev, and A.~Polkovnikov.
\newblock {\em Phys. Rev. B}, 81:012303, 2010.

\bibitem{albu10}
A.~F. Albuquerque, F.~Alet, C.~Sire, and S.~Capponi.
\newblock {\em Phys. Rev. B}, 81:064418, 2010.

\bibitem{mukherjee11}
V.~Mukherjee, A.~Polkovnikov, and A.~Dutta.
\newblock {\em Phys. Rev. B}, 83:075118, 2011.

\bibitem{david09}
D.~Schwandt, F.~Alet, and S.~Capponi.
\newblock {\em Phys. Rev. Lett.}, 103:170501, 2009.

\bibitem{rams11}
M.~M. Rams and B.~Damski.
\newblock {\em Phys. Rev. Lett.}, 106:055701, 2011.

\bibitem{thakurathi12}
M.~Thakurathi, Diptiman Sen, and A.~Dutta.
\newblock {\em Phys. Rev. B}, 86:245424, 2012.

\bibitem{patel13}
A.A. Patel, S.~Sharma, and A.~Dutta.
\newblock {\em Europhys. Lett.}, 102:46001, 2013.

\bibitem{chen08}
Shu Chen, Li~Wang, Yajiang Hao, and Yupeng Wang.
\newblock {\em Phys. Rev. A}, 77:032111, 2008.

\bibitem{divakaran13}
U.~Divakaran.
\newblock {\em Phys. Rev. E}, 88:052122, 2013.

\bibitem{quan06}
H.~T. Quan, Z.~Song, X.~F. Liu, P.~Zanardi, and C.~P. Sun.
\newblock {\em Phys. Rev. Lett.}, 96:140604, 2006.

\bibitem{rossini07}
D.~Rossini, T.~Calarco, V.~Giovannetti, S.~Montangero, and R.~Fazio.
\newblock {\em Phys. Rev. A.}, 75:032333, 2007.

\bibitem{damski11}
Bogdan Damski, H.~T. Quan, and Wojciech~H. Zurek.
\newblock {\em Phys. Rev. A}, 83:062104, 2011.

\bibitem{mukherjee12}
V.~Mukherjee, S.~Sharma, and A.~Dutta.
\newblock {\em Phys. Rev. B}, 86 (R):020301, 2012.

\bibitem{nag12}
T.~Nag, U.~Divakaran, and A.~Dutta.
\newblock {\em Phys. Rev. B}, 86 (R):020401, 2012.

\bibitem{sharma12}
S.~Sharma, V.~Mukherjee, and A.~Dutta.
\newblock {\em Eur. Phys. J. B}, 85:143, 2012.

\bibitem{sharma121}
S.~Sharma and A.~Rajak.
\newblock {\em J. Stat. Mech}, page P08005, 2012.

\bibitem{peres84}
A.~Peres.
\newblock {\em Phys. Rev. A}, 30:1610, 1984.

\bibitem{peres93}
A.~Peres.
\newblock {\em Quantum Theory: Concepts and Methods}.
\newblock Kluwer Academic Publishers Dordrecht, 1993.

\bibitem{jalabert01}
R.~A. Jalabert and H.~M. Pastawski.
\newblock {\em Phys. Rev. Lett.}, 86:246, 2001.

\bibitem{karku02}
Z.~P. Karkuszewski, C.~Jarzynski, and H.~W. Zurek.
\newblock {\em Phys. Rev. Lett}, 89:170405, 2002.

\bibitem{cerruti02}
N.~R. Cerruti and S.~Tomsovic.
\newblock {\em Phys. Rev. Lett}, 88:054103, 2002.

\bibitem{cucchietti03}
F.~M. Cucchietti, D.~A.~R. Dalvit, J.~P. Paz, and W.~H. Zurek.
\newblock {\em Phys. Rev. A}, 95:105701, 2003.

\bibitem{zheng08}
Q.~Zheng, W.~G. Wang, X.~P. Zhang, and Z.~Z. Ren.
\newblock {\em Phys. Lett. A}, 372:5139, 2008.

\bibitem{zheng09}
Q.~Zheng, W.~G. Wang, P.~Q. Qin, P.~Wang, X.~P. Zhang, and Z.~Z. Ren.
\newblock {\em Phys. Rev. E}, 80:016214, 2009.

\bibitem{huan09}
J.~F. Huan, Y.~Li, J.~Q. Liao, L.~M. Kuang, and C.~P. Sun.
\newblock {\em Phys. Rev. A}, 80:063829, 2009.

\bibitem{buchkremer00}
F.~B.~J. Buchkremer, R.~Dumke, H.~Levsen, G.~Birkl, and W.~Ertmer.
\newblock {\em Phys. Rev. Lett.}, 85:3121, 2000.

\bibitem{zhang09}
J.~Zhang et.al.
\newblock {\em Phys. Rev. A}, 80:012305, 2009.

\bibitem{sanchez09}
C.~M. S\'anchez, P.~R. Levstein, R.~H. Acosta, and A.~K. Chattah.
\newblock {\em Phys. Rev. A}, 80:012328, 2009.

\bibitem{venuti07}
L.~C. Venuti and P.~Zanardi.
\newblock {\em Phys. Rev. Lett}, 99:095701, 2007.

\bibitem{kopp05}
A.~Kopp and S.~Chakravarty.
\newblock {\em Nat. Phys}, 1:53, 2005.

\bibitem{divakaran07}
U.~Divakaran and A.~Dutta.
\newblock {\em J. Stat. Mech}, page P11001, 2007.

\bibitem{lieb61}
E.~Lieb, T.~Schultz, and D.~Mattis.
\newblock {\em Ann. Phys.,NY}, 16:37004, 1961.

\bibitem{pfeuty70}
P.~Pfeuty.
\newblock {\em Ann. Phys. (NY)}, 57:79, 1970.

\bibitem{bunder99}
J.~E. Bunder and R.~H. McKenzie.
\newblock {\em Phys. Rev. B.}, 60:344, 1999.

\bibitem{damski13a}
Bogdan Damski.
\newblock {\em Phys. Rev. E}, 87:052131, 2013.

\bibitem{damski13b}
B~Damski and M.~M. Rams.
\newblock arxiv:1308.5917, 2013.

\bibitem{zanardi06}
P.~Zanardi and N.~Paunkovi{\'c}.
\newblock {\em Phys. Rev. E}, 74:031123, 2006.

\bibitem{yang08}
Shou Yang, Shi-Jian Gu, Chang-Pu Sun, and Hai-Qing Lin.
\newblock {\em Phys. Rev. A}, 78:012304, 2008.

\bibitem{gu08}
Shi-Jian Gu, Ho-Man Kwok, Wen-Qiang Ning, and Hai-Qing Lin.
\newblock {\em Phys. Rev. B}, 77:245109, 2008.

\bibitem{deng09}
S.~Deng, G.~Ortiz, and L.~Viola.
\newblock {\em Phys. Rev. B}, 80:241109, 2009.

\bibitem{mukherjee10}
V.~Mukherjee and A.~Dutta.
\newblock {\em Europhys. Lett.}, 92:37004, 2010.

\end{thebibliography}

\end{document}